# Cost of Implementation of Basel III reforms in Bangladesh: A Panel data analysis

Dipti Rani Hazra[1]
Md. Shah Naoaj[2]
Mohammed MahinurAlam[3]
Abdul Kader[4]

***Abstract:***
*Inspired by the recent debate on the macroeconomic implications of the new bank regulatory standards known as Basel III, we tried to find out in this study that the impact of Basel III liquidity and capital requirements in Bangladesh proposed by Basel Committee on Banking Supervision (BCBS, 2010a). A small set of macro-variables, using a sample of 22 private commercial banks operating in Bangladesh for the period of 2010-2014, are used to estimate long-run relationships among the variables. The macroeconomic variables are included The profitability of banks, GDP, banks' lending to private sector, Net Stable Funding Ratio, Tier 1 capital Ratio, Interest rate spread, real interest rate. The cost is quantified using Driscoll and Kraay panel data models with fixed effect. Impact of higher capital and liquidity requirement on Interest rate spread and lending to private sector of banks were considered as the cost to the economy as a whole whereas impact of higher capital and liquidity requirement on profitability of banks(ROE) was considered as the cost of banks. Here it is found that, the interest rate level is positively affected by the tighter liquidity and capital requirements which driven toward lessen of the private sector lending of banks. The return on equity of banks varies negatively with the liquidity and capital. The economic costs are considerably below the estimated positive benefit that the reform should have by reducing the probability of banking crises and the associated banking losses (BCBS, 2010b).*

[1] Joint Director, Basel Framework Implementation Section, BRPD, Bangladesh Bank.
[2] Assistant Director, Basel Framework Implementation Section, BRPD, Bangladesh Bank.
[3] Joint Director, Chief Economist Unit, Bangladesh Bank.
[4] Senior Officer, One Bank Ltd.

# 1 Introduction

Recent global financial crisis comparatively didn't touch Bangladesh, its economy and banking sector proved risk resilient in this regard while the south East Asian region could not avoid the spillover effects of the subsequent global financial crisis. The global regulatory response to the financial crisis is Basel III reforms which complements the Basel II and Basel I frameworks, offers a valuable opportunity to make banking system more risk resilient. The main spirit of Basel III reforms is the more risk resilient banks and banking industry with higher quality capital requirement, liquidity standards and cap on both on and off balance sheet leverage which are required to implement by 2018 in phased in manner(BCBS, 2010d). Basel Committee on Banking Supervision (BCBS, 2010b) summarizes the economic effects analysis named the report of LEI (Long term Economic Impact) of the proposed liquidity and capital regulation on output. This paper details one of the many pieces of work. The following issue is tried to be explored carefully in this paper: what is the impact of higher capital requirements and tighter liquidity regulation on the long-term economic performance of the Bangladesh?

# 2 Literature Review

An increase in the bank intermediation cost is the route through which the economic activity is influenced by the changes in the regulations of liquidity and capital. For compensating the holding cost of more liquidity and capital, the lending rates will be increased by the banks which cause to increase the banks' interest rate spread. Lower output and lower investment will be occurred due to the unsatisfactory substitutability of different market financing forms and banks credit.

There are different macroeconomic models which can be used as the basement for the calculation of the economic costs of the higher liquidity and capital requirements for the output level. In many aspects, the model differs. First they refer different areas or different countries, second, based on full estimation, some are calculated, and on the other hand, others are totally standardized (on the light of particular parameters, generally the coefficients values are generated from studies casting, microeconomic or unrelated sectors). Finally, the bank's role for liquidity and capital and the banking sector are explicitly featured by some models whereas others generally don't.

For calculating long term output reduction which is occurred by spreads of higher lending creating from stronger liquidity and capital standards, an error correction model is used by Wong et al. (2010) and Gambacorta (2010). In the steady state, assistance to disentangle loan supply and loan demand factors is the key benefit of this approach. On the basis of cumulated historical data, the long-run relationship between the output reduction and liquidity or capital can be established. Disallowance of inauguration of countercyclical capital buffers and other similar counter-factual experiment conduct is the prime demerits of this approach (BCBS, 2010b).

The costs for the crisis of a Euro area from both higher liquidity and capital applying a DSGE model which includes a banking sector and financial frictions are calibrated by Roger et al. (2010). Explicitly the credit markets and balance sheet of banks are featured by their paper. For analyzing the way of changing impact of liquidity and capital requirements on the banking output and conditions (lending and spreads), a unified framework is provided by it. In a conceptually steady manner, the experiments of counter-factual policy are allowed by the DSGE models. On the other hand, the estimation process if often intimidating due to its totally calibrated nature. Van den Heuvel (2008) cited the DSGE models where still experimental position is hold by Meh and Moran (2008) which indicates the disintegration in their process of making policy.

Semi-structural models are used by Locarno (2004); conversely as input variables, the income statements and balance sheet conditions of banks are not directly incorporated by these models. Instead, lending spreads and other similar variables must be incorporated as the effects of these models. Mapping the effect of the higher liquidity and capital requirements on spreads of lending is the first step according to this idea. However, estimating the effect of the liquidity or capital cost on output is so much difficult task. Besides, because of the models' size, it is also very difficult to compute the long-term effects and through over a convincingly huge number of years, simulations can approximately compute the cost.

An alternative channel is assumed by Miles et al. (2011) through which the economic activity can be influenced by the changes in capital, which can be occurred via the rise of the bank intermediation cost of funding. In terms of external financing, the customers can be suffered by the increase of funding costs from banks (WACC) which will be passed to banks primarily and ultimately will be passed to the customers of the banks. Single standard production function is assumed as a way of generating output of production by firms' labor and capital.

For all Public Commercial Banks (PCBs) operated in Bangladesh especially for newly introduced PCBS, the capital ratio is significant and positive to the interest rate spread (Monjur, 2010). According to this, it is clear that these banks' capital bases are the high bank earnings which are generated from the high margins. Economies of scale are also another base for PCBs in Bangladesh which is indicated by the significant and negative impact on interest margin by the size of the banks.

The higher capital and liquidity requirements of banks will change the steady-state output. Consequently, for estimating the banks' higher WACC's loss of calibrated output, a function of production with a standardized elasticity of substitution can be applied by them.

# 3 Basel III Implementation Status in Bangladesh

From January 2015, Basel III has started to be implemented by Bangladesh Bank in Bangladesh. The capital adequacy requirements are needed to meet by the banks which is much higher than global capital standards' Basel III in accordance with the 18/2014 BRPD circular of Bangladesh Bank on the guidelines of Risk Based Capital Adequacy. Total CAR of 10%, Tier 1 CAR of 6% and Tier 1 (CET1) CAR of 4.5% is required to meet by the banks from 1st January, 2015. Compared to the minimum requirements of the Basel III of 8%, 6% and 4.5% respectively for Total CAR, Total Tier 1 and CET1, these standards are much higher. 10% Total CAR is remained unchanged according to the current requirements of BB. Above the minimum requirement of capital adequacy, 2.5% buffer of capital conservation is introduced by BB aligning with the requirements of Basel III. Banks are required to keep this buffer in phased in manner (0.625% annual increment).The banks will meet it along with the CET1 capital. A CET1of 7% will be required to be met along with the buffer of capital conservation.

BB definition of capital (BRPD 18/2014), the risk coverage (credit, market and operational risks as set out in chapter 05, 06 and 07 respectively) and the method of calculation are compliant with the requirements set out in the applicable Basel standards. For implementing the Basel III in Bangladesh, transitional arrangements are issued by BB according to the guidelines of Basel Framework. In the following section, the phase wise arrangements for the implementation of Basel III in Bangladesh are given:

**Table 1: Phase-in arrangements for Basel III implementation in Bangladesh[5]**

| | 2015 | 2016 | 2017 | 2018 | 2019 |
|---|---|---|---|---|---|
| Minimum Common Equity Tier-1 (CET-1) Capital Ratio | 4.50% | 4.50% | 4.50% | 4.50% | 4.50% |
| Capital Conservation Buffer | - | 0.625% | 1.25% | 1.875% | 2.50% |
| Minimum CET-1 plus Capital Conservation Buffer | 4. 50% | 5.125% | 5.75% | 6.375% | 7.00% |
| Minimum T-1 Capital Ratio | 5.50% | 5.50% | 6.00% | 6.00% | 6.00% |
| Minimum Total Capital Ratio | 10.00% | 10.00% | 10.00% | 10.00% | 10.00% |
| Minimum Total Capital plus Capital Conservation Buffer | 10.00% | 10.625% | 11.25% | 11.875% | 12.50% |
| Phase-in of deductions from CET1 | | | | | |
| Excess Investment over 10% of a bank's equity in the equity of banking, financial and insurance entities | 20% | 40% | 60% | 80% | 100% |
| Phase-in of deductions from Tier 2 Revaluation Reserves (RR)[6] | | | | | |
| RR for Fixed Assets, Securities and Equity Securities | 20% | 40% | 60% | 80% | 100% |
| Leverage Ratio | 3% | 3% | 3% Readjustment | Migration to Pillar 1 | |
| Liquidity Coverage Ratio | ≥100% (From Sep.) | ≥100% | ≥100% | ≥100% | ≥100% |
| Net Stable Funding Ratio | > 100% (From Sep.) | >100% | >100% | >100% | >100% |

# 4 Data description

## 4.1 Definitions of Capital and Liquidity

Before accomplishing the empirical work, it is quite important to define the liquidity and capital clearly. The bank capital is represented by the variable of Tier 1 capital ratio[7]. Similarly, the Loans to-Deposits ratio is the most widely followed liquidity variable. Nevertheless, different ratios are focused under Basel III. Since the banking capital's highest quality component is the tangible common equity, for the base of the capital, tangible common equity (TCE) capital is required to be focused more in Basel III which is the first breakthrough to TCE/RWA (risk-weighted assets).

$$\text{TCE/RWA} = \frac{\text{Common equity} - \text{Intangibles } - \text{goodwill}}{\text{Risk Weighted Assets}}$$

With the application of NSFR[8,9] (net stable funding ratio) which can be formulated by ASF (available amount of stable funding) divided by RSF (required amount of stable funding), the long-term adequacy banks' liquidity can be evaluated by the Basel III. Debt, equity and other liabilities are included in the ASF with a 1 year effective maturity or the larger, stable deposits of 85% with less than 1 year residual maturity, and less stable deposits of 70% with less than 1 year residual maturity. Govt Debt (government debt) of 5%, Other Assets of 100% (excluding interbank loans and cash), Ret Loans (Retail loans) of 85% having less than 1 year maturity and Corp Loans (Corporate loans) of 50% are included in the RSF. Accordingly,

$$\text{NSFR} = \frac{\text{CEquity} + \text{Debt_1yr} + \text{Liabs_1yr} + 85\%\text{StbDeposits} < 1yr + 70\% OtherDeposits < 1yr}{5\%\text{GovtDebt} + 50\%\text{CorpLoans} < 1yr + 85\% RetLoans < 1yr + 100\%\ OtherAssets}$$

[5]Bangladesh Bank(2015), "Guidelines on Risk Based Capital Adequacy (in line with Basel III capital requirements)"
[7] With the denominator comprising the RWA (risk weighted assets), in the numerator on a basis of `going concern', a capacity of loss-absorbing is possessed which includes other qualifying financial instruments plus common equity in the entire Tier 1 capital ratio.
[8]For assessing the banks' liquidity adequacy in the short run (Say 30 days), LCR (liquidity coverage ratio) is used by the BCBS (2010a).
[9]For allowing the comparison with its impact on other studies, NSFR December 2009 definition have been used here (BCBS, 2009).

The Net Stable Funding Ratio should be more than ONErequired by the Basel III which indicates that the uses of the funding are lower than the funding sources. For assessing the effect of liquidity requirements of the Basel III through transferring NSFR into the ratio of Loans-to-Deposits, the similar approach has been used in previous studies. In the two ratios, 46 basis points will be decreased by the correspondent increase of 1% increase in the NSFR according to Wong et al. (2010). On average, it is still very hard to estimate the linkage between the ratio of Loans-to-Deposits and NSFR since it is assumed that the Loans-to-Deposits ratio and NSFR have a linear relationship and with a tiny data sample. In our study we calculate the NSFR of 26 scheduled banks of Bangladesh for 2010-2014 through reviewing the historical income statements and balance sheets.

## 4.2 Data Description

On the basis of specific yearly data of bank from 2010-2014, the analysis of the panel data is conducted with following variables:
(i) the logarithm of GDP at current market price[10](Y); (ii) the spread between the short-run rate of interest and average lending rate $(r - i)$; (iii) logarithm of private sector's lending(L); (iv) logarithm of banks Return on equity's (ROE); (v)logarithm of liquidity-to-deposit ratio's (LIQ); (vi)logarithm of TIER1/RWA (CAP); (vii) Logarithm of Lending to GDP ratio. During the detection of seasonality, seasonal adjustment is completed of all these series.

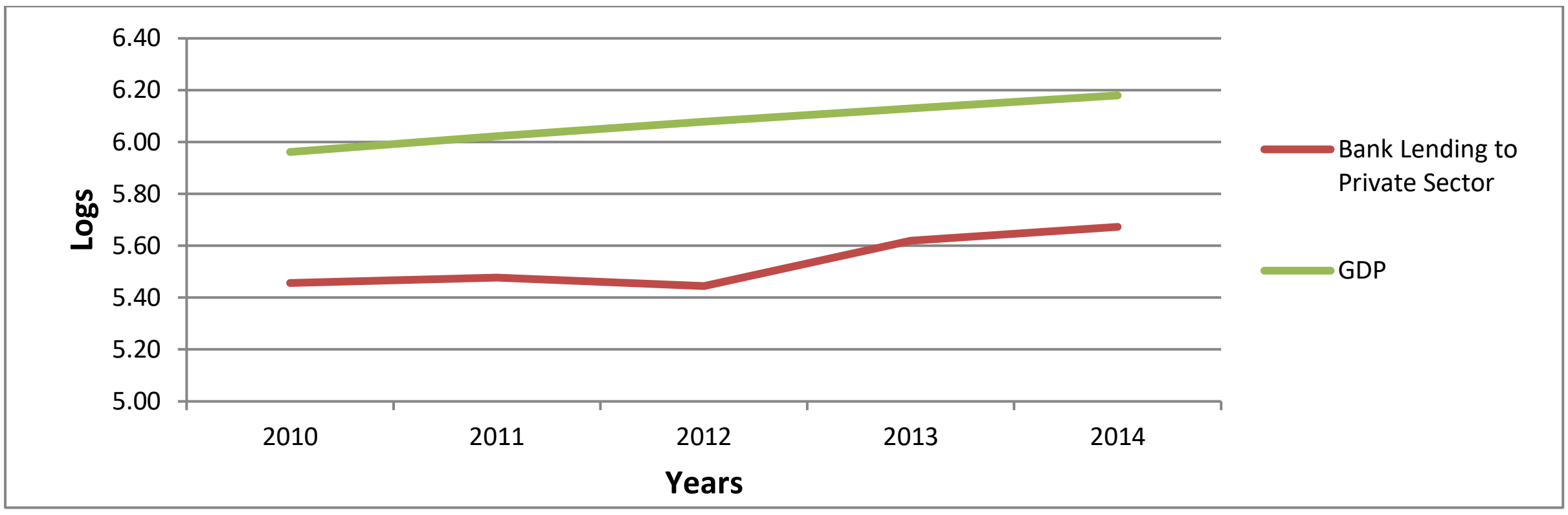

Figure 01: Bank Lending to Private Sector and GDP

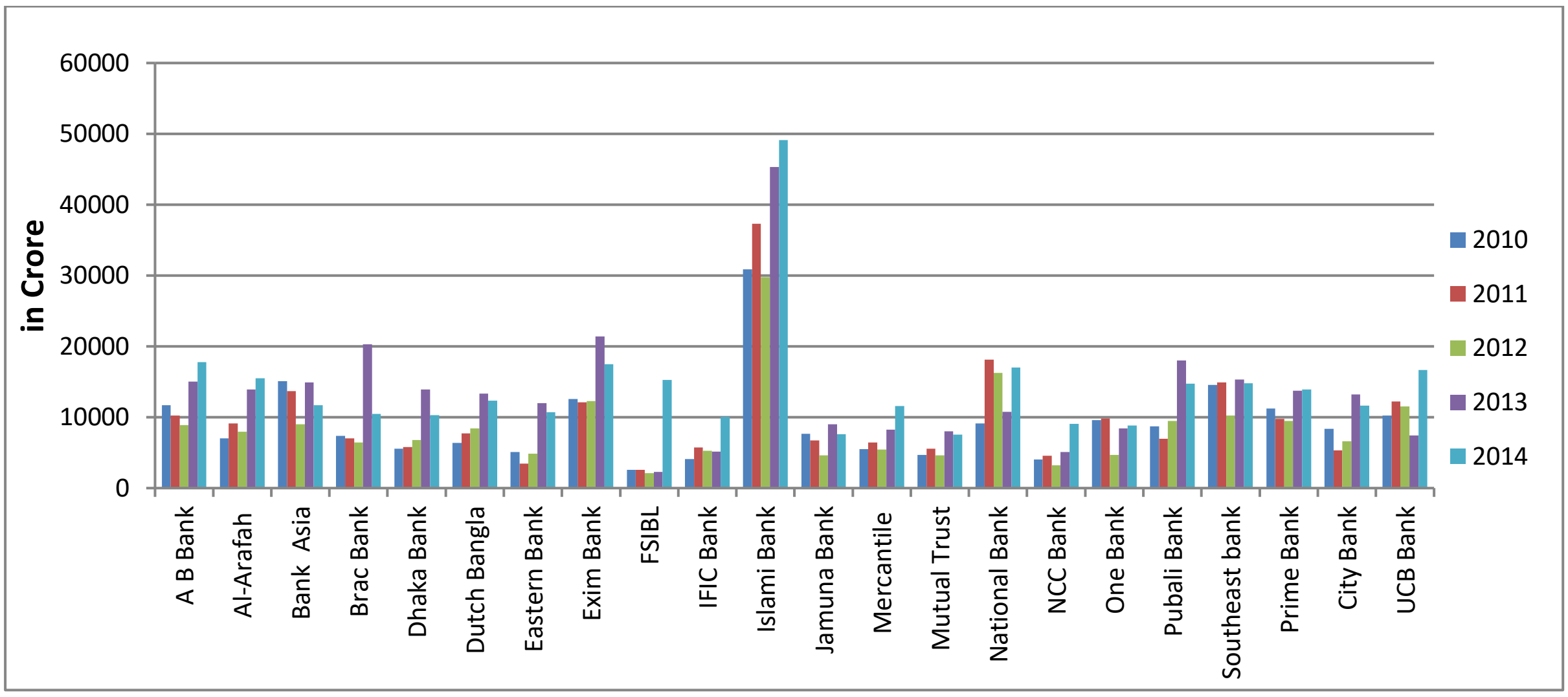

[10] Data Source: http://www.mof.gov.bd/en/budget/14_15/ber/en/Ch-02%20%28English-2014%29_Final_Draft.pdf

Figure 02: Bank specific lending to private sector[11]

In figure 1, it is reported to show the private sector lending of the banks and annual GDP behavior by a graphical analysis where in figure 2, it is shown the bank specific private sector lending. Between the series (except political instability affected during 2012), a strong correlation is represented here, which suggests integrated relationship among them. For instance, the number of projects is generally increased by the improved economic conditions which in terms of anticipated net present value become profitable and it ultimately raises the credit demand (Kashyap et al., 1993). It refers a long term as well as strong relationship between the GDP and the Credit.

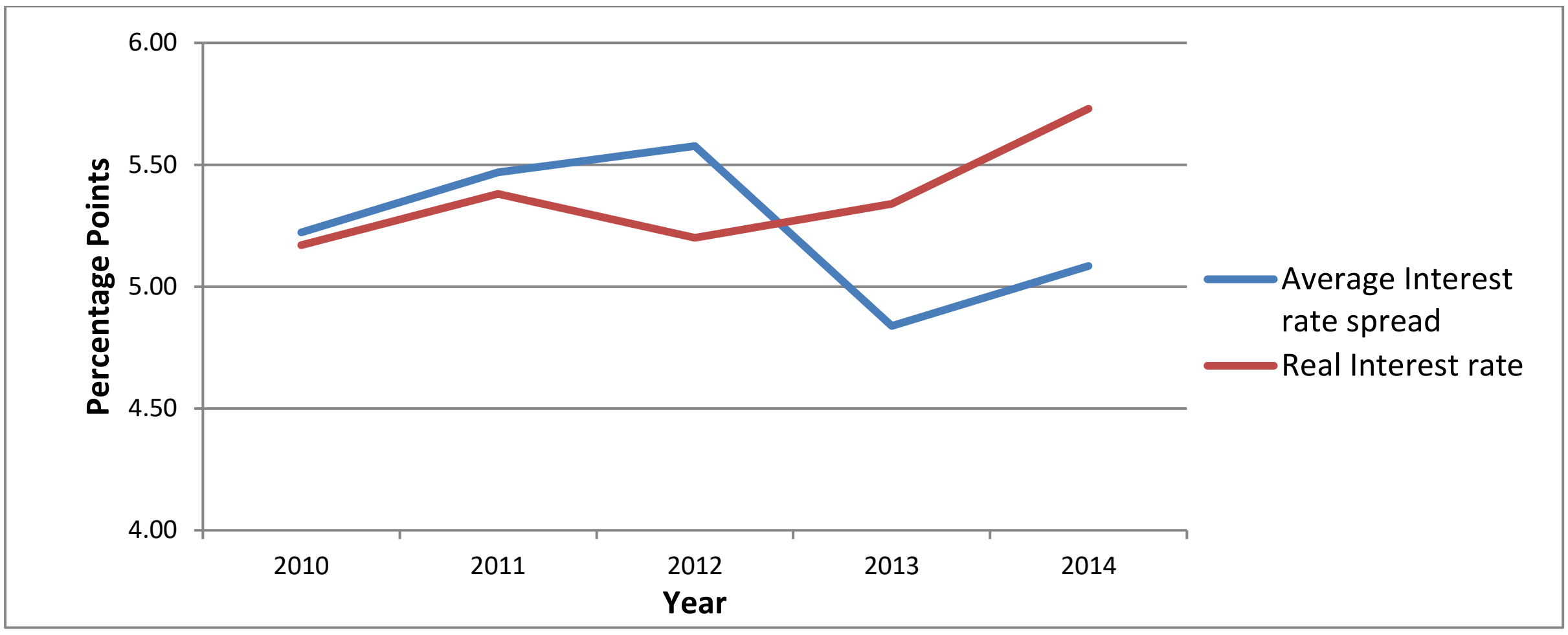

Figure 03: Bank interest rate spread and real interest rate

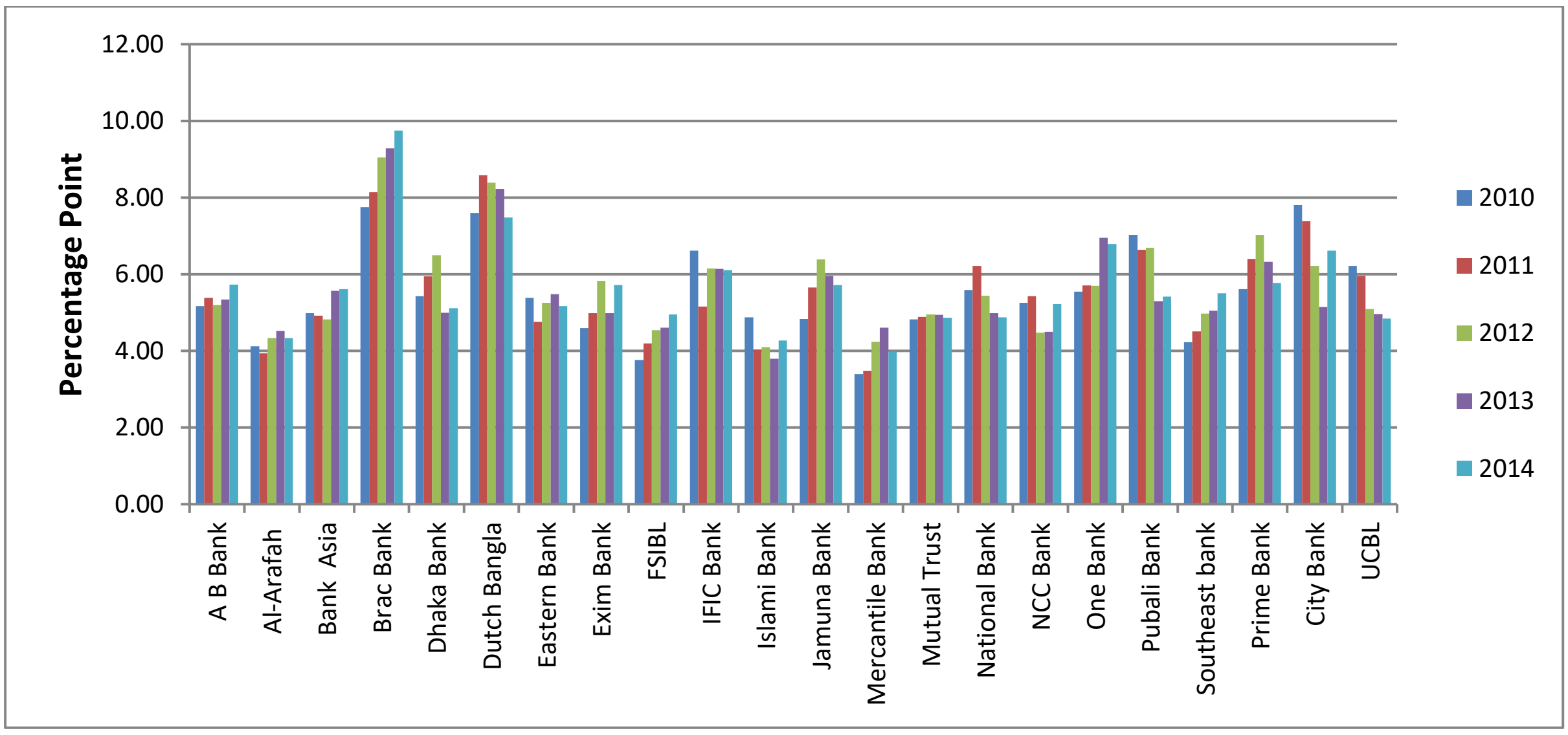

Figure 04: Bank specific Interest Rate Spread

On diverse types of loans like industrial and commercial lending rate and mortgage rates, weighing bank lending rates obtain the composite arte of lending (r). The consequent loan category gives relative importance to the weights. Average 12 months of interbank interest rate minus CPI inflation gives the real interest rate $(i - \pi)$. In figure 4 the real interest rate and the spread behavior $(r - i)$ are reported. In Bangladesh, due to political instability in the year 2012–2013, the spread of banks fell. Conversely, throughout these times, an increasing trend in the real interest rate is found.

[11] Data Source: Statistics Department (Banking Statistics Division), Bangladesh Bank.

Due to financial innovation and turmoil of financial sector because of political instability as well as capital market debacle, the banking industry enters into excess liquidity regime. The highest deposits risk attract more deposits to banking industry while banks failed to invest those moneys in the form of loans and investment which cause the arising of excess liquidity scenario. In 2012, the liquidity scenario was at the pinnacle and the same time the ROE comes to the lowest point which shows the negative relationship between liquidity and profitability. At the same time the capital position of banking industry increases steadily throughout these periods which happened because of Basel II requirements implemented these periods. According to the expectation, at the time of financial turmoil, a significant drop is exhibited by the bank profitability.

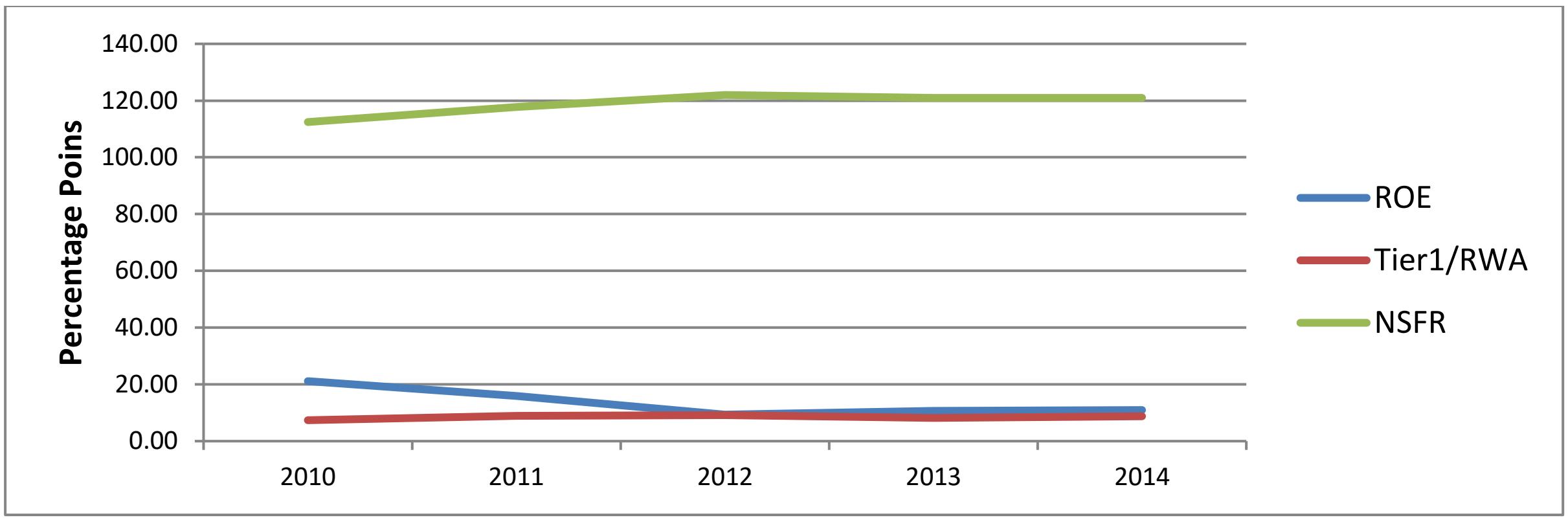

Figure 05: Bank ROE, Tier1/RWA and NSFR

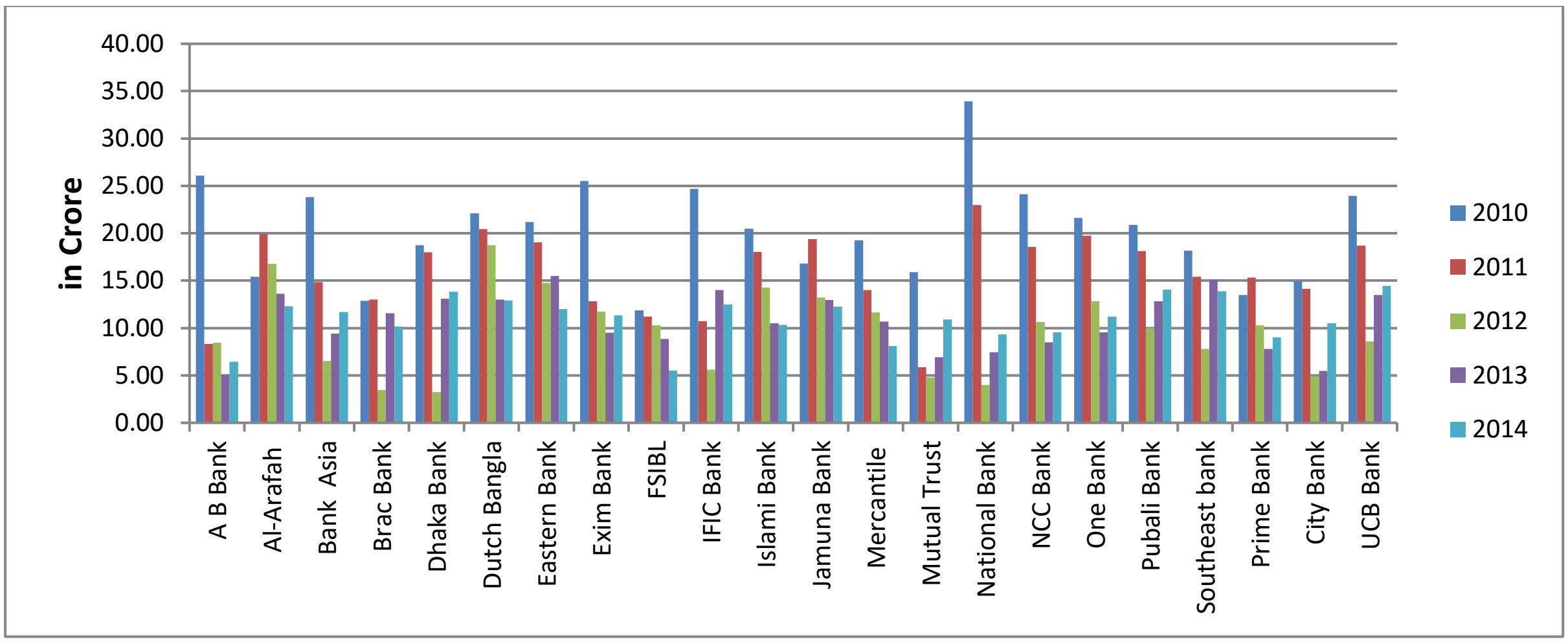

Figure 06: Bank specific ROE

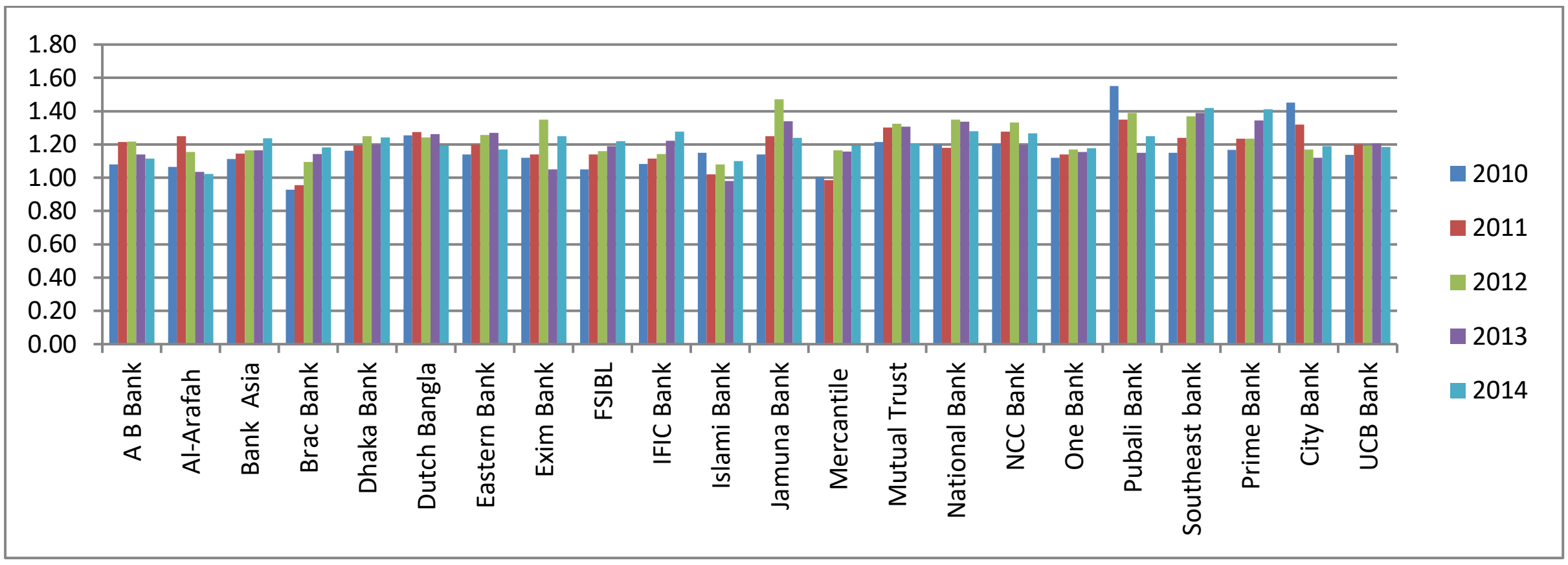

Figure 07: Bank specific NSFR[12]

# 5 Methodology:

## 5.1 The sample

The sample consists of 22scheduled private commercial banks for a period of five years (2010-2014). The choice of the 22 scheduled banks is not random since in Bangladesh, state-owned government banks and specialized banks are formed for some special purposes and they are not regulatory compliance in capital adequacy we did not consider those banks in our analysis. The bank-specific data were collected from Bangladesh Bank Statistics Department, Department of Offsite Supervision and banks' annual reports. The macroeconomic data were collected from the Ministry of Finance database and Bangladesh Bureau of Statistics database.

## 5.2 The model

Because of limited data availability and because of the cross-country nature of our study, we adopt a panel data model to estimate the long-run impact of changes in capital buffers.

Several methods and approaches can be used to estimate the cost of higher capital and liquidity standards proposed in Basel III reforms. Generally time series models are used to estimate the long run impact of these variables and DSGE models (see Christiano et al, 2010) (see, among others, Gambacorta 2011). Since Basel II standard implemented in Bangladesh from year 2010 we don't have data related to capital adequacy prior to 2010 hence we use a panel data model to estimates the cost of implementation of Basel III reforms in Bangladesh. The literature of BCBS (2010b) and Gambacorta (2010) is followed for estimating both higher bank liquidity and capital requirements' long-term cost effect and for estimating the long-run connection between a tiny set of variable for the country, Regression Model of Panel Data is used here. The main focus of this analysis is given on the long-term impacts on bank profitability, GDP, lending and interest rates of the requirements of the Basel III.It establishes a framework to estimate the effects of higher bank capital and liquidity on output and bank profitability.

Interest rate spread equation is the first relation. Rochet (1997) show that in a model of imperfect competition among N banks each one sets the lending rate as the sum of the exogenous cost of banks' refinancing on the money market, other costs (such as bank capital and liquidity requirements) and a constant markup. Economic theory therefore shows that the difference between the lending rate and the money market rate can be represented as:

$$(r - i)_{it} = \gamma_0 + \gamma_1 LIQ_{it} + \gamma_2 CAP_{it} + \varepsilon_{it} \quad \text{(i).}$$

For provided liquidity and capital levels, a loan supply which is horizontally perfect can also be interpreted by this equation.

The equation of lending demand is the second relationship. Bank lending demand should be a positive function of real GDP and a negative function of the spread, the following type long-run log-linear relationship is also supposed to be existed.

$$L_{it} = \beta_0 + \beta_1 GDP_{it} + \beta_2 (r - i)_{it} + \varepsilon_{it} \quad \text{(ii).}$$

The third relationship is the reduction form of the bank's profit. The financial deepening and the spread (the ratio of total lending and GDP) critically influence the banks' the long-term return on equity. We have:

$$ROE_{it} = \delta_0 + \delta_1 L_{it} + \gamma_2 (r - i) + \varepsilon_{it} \quad \text{(iii).}$$

Where,

[12] NSFR is writers own calculation

CAPit is the common equity Tier 1 capital to risk-weighted assets i at time t.
LIQit is net stable funding ratio calculated based on the definition in the Dec 2009 proposal i at time t.
(r− i)it is the interest rate spread i at time t.
Lit is the ratio of lending to private sector to GDP i at time t.
GDPitis the Nominal GDP of Bangladesh i at time t.
ROEitis the Return of Equity of i at time t.

## 6 Analysis of Results

The Harris–Tzavalis test is used for sorting out the Unit root test in this study and on Table 1 the results are presented. At 0.05 level of significance, the values of pare significant indicated by the results hence the null hypothesis is not accepted here and conclude in such a way that the data is immobile where differencing is not required. Asymptotically, the panels' number ratio to time periods to zero is required due to the test of Levin–Lin–Chu and these data sets are not well suited along with a huge panels' number and few timeframe relatively. Harris–Tzavalis test has been used here which presumes that the panels' number likely to infinity where fixed timeframe is given.

Table: 01

| Variables name | Rho | P value |
|---|---|---|
| CAP | -0.0196 | 0.0000 |
| r − i | 0.2785 | 0.0165 |
| LIQ | 0.2020 | 0.0021 |
| GDP | 0.0116 | 0.0000 |
| L | 0.2011 | 0.0020 |
| ROE | 0.1175 | 0.001 |

The Harris–Tzavalis test is significant at all the usual testing levels. Therefore, we reject the null hypothesis and conclude that all the variables are stationary.
Using the specification simple within the groups, estimation will be conducted. Panel nature and Data set's exigencies are the simple reasons behind the selection of Fixed Effect specification. In addition, standard errors of Driscoll and Kraay (1998) will also be applied to eradicate the group biasness and make data more accurate to account for correlations, autocorrelation and heteroskedasticity between groups. In each case, a unit root's null hypothesis was rejected convincingly.
The following calculation shows the projected long-term relationships:

$$(r - i)_{it} = 1.617 + \underset{(0.00)}{0.639}\, LIQ_{it} + \underset{(0.00)}{0.169}\, CAP_{it} \qquad \text{(iii)}$$

$$L_{it} = 3.29 + \underset{(0.03)}{1.352}\, GDP_{it} - \underset{(0.05)}{0.306}\, (r - i)_{it} \qquad \text{(iv)}$$

$$ROE_{it} = \delta 0 + \underset{(0.09)}{1.36}\, (L/GDP)_{it} - \underset{(0.10)}{1.06}\, LIQ_{it} - \underset{(0.03)}{0.49}\, CAP_{it} \qquad \text{(v)}$$

## 7 Discussion of Findings

With a 5% P-value, the over-identified restrictions set are established. As for the estimated coefficients, the long-run elasticity between the spread and the two regulatory variables are statistically significant at 0% level. But the coefficient is largely sensitive to the change in the interest rate spread rather than Tier 1 capital. In case of 1 percentage point increase of the liquidity ratio the spread increases by 0.639 percent whereas for 1 percentage point increases in Tier 1capital the spread increases by 0.169 percent. The relationship between capital and spread is similar to the findings of Gambacorta (2011), MAG (2010) and BCBS (2010b).The significant relationship between capital level and interest rate spread found in study on Private Commercial Banks in Bangladesh conducted by Monzur (2010). For all PCBs especially for the newly started PCBs, significant and positive capital ratio is present to interest margins. These banks capital base are channeled by the contribution of high margins to the high earnings of the banks. GDP and private sector lending's long-run elasticity is 1.352 that is in line with the euro area's result (Calzaet al., 2006; Gambacorta and Rossi, 2010). Some variables omission from the model like house or wealth purchases that aren't taken over by GDP transactions is likely to be reflected by the above income elasticity.

The elasticity of bank lending to private sector with respect to the spread is negative (−0.306).That means because of increasing capital and liquidity standards increase the interest rate spread which causes decrease of banking sectors lending to private sectors.

As for the third model, the elasticity between ROE and L/Y is quite high (1.36).The exclusion of several variables from the model like behavior of the stock market through which non-interest income element dynamic could be captured are the reasons behind the elasticity. Respectively negative (0.49) and (1.06) is found to bank loans elasticity with Tier 1 Capital and the Liquidity. As Basel III standards increase the capital requirement which is 12.5 %(including capital conservation buffer) the lending to private sector is expected to fall because of increasing interest rate spread. Increased capital and liquidity standard together with decreased lending to private sector and increased interest rate spread will reduce the profitability of the banking sector in Bangladesh.

## 8 Policy Implications

This study has found negative relationship between enhanced liquidity and capital requirement on banks' lending to private sector. The negative impact of liquidity on lending is higher than the negative impact of higher capital conservation requirement. Moreover, the study found negative impact of enhanced liquidity and Tier 1 capital on profitability of Banks. Since, the study found significant cost of Basel III implementation on Banks, it can be said that conserving higher capital and liquidity is not the only remedy for preventing banking debacle. Therefore, following steps can be taken for implementation of Basel III and for preventing banking collapse- a) Basel III implementation time frame can be stretched to 2021 instead of 2019 to give more times to Banks to cope up with the new requirements, b) timely and high quality disclosure should be ensured by the Central Bank through enhanced supervision, c) sound corporate governance needs to be ensured for preventing undue and aggressive lending, d) a crisis management system should be introduced, e) suitable legal and institutional framework should be implemented for preventing banking crisis, f) quality supervision from central bank should be ensured for preventing plundering of Banks' money through aggressive lending, g) improving infrastructure and ensuring uninterrupted power and energy supply for creating opportunity of lending. By taking above measures the negative impact of Basel III implementation can be mitigated and a sound banking system can be established.

## 9 Summary and Conclusions

In response to new regulatory standards' (Basel III reforms),we estimate the long-term economic costs of implementations of Basel III in Bangladesh using a small macro-variables set on 22 Private Commercial Banks over the timeframe of 2010 to 2014 applying a Panel Data Regression models. It shows that tighter capital and liquidity requirements have negative (but rather limited) effects on the level of output and more sizeable effects on bank profitability. Each 1 percentage point increase of the liquidity ratio the interest rate spread increases by 0.639 percent whereas for 1 percentage point increases in Tier 1capital ratio increases the interest rate spread by 0.169 percent. These results are in line with other studies (MAG, 2010b; Angelini*et al.*, 2011, Monzur, 2010).

The elasticity of bank lending to private sector with respect to the spread is negative (−0.306). That means because of increasing capital and liquidity standards increase the interest rate spread which causes decrease of banking sectors lending to private sectors. Finally, As Basel III standards increase the capital requirement which is 12.5 %(including capital conservation buffer) the lending to private sector is expected to fall because of increasing interest rate spread. Increased capital and liquidity standard together with decreased lending to private sector will reduce the profitability of the banking sector in Bangladesh. Due to ignorance of the major recession in the analyzed period here and failure to estimate the macro impact in the long run, it is very essential to scrutinize the result of this study very carefully.

Overall, in comparison with the positive benefits got from the tighter liquidity and capital standards such as reduction of banking losses and possibility of occurring banking crises, the associated economic cost for such tighter standards are lower (BCBS, 2010a, b). However, uncertainty is present in the reduction of the possibility of crises and on the other hand. For the economy and the bank behavior at large, Basel III reforms package implications can't be captured by any single approach and for further and clearer realization of the key connection at work, more research work on this area is required. There are several limitations that are faced during the preparation of this paper. First, since 1971, there is no evidence of such banking crisis in Bangladesh which restrict in evaluating the net economic advantages and probable economic advantages taking into account of the output losses incurred in Bangladesh from the requirements of the Basel III. The assumptions followed for calculating the NSFR are sensitive to the NSFR requirements of the estimated costs. For quantifying the long-term economic effect of the latest agreed changes to the liquidity and capital international standards for the banking sector of the country on its economy, we have tried hard notwithstanding this. For the economy and the bank behavior at large, Basel III package implications can't be captured by any single approach and for further and clearer realization of the key connection at work, more research work on this area is required.